\documentclass[preprint, 12pt,3p,lefttitle]{elsarticle}

\usepackage{xcolor}
\usepackage[utf8]{inputenc} 
\usepackage[T1]{fontenc}  
\usepackage{ragged2e}
\usepackage{amssymb}
\usepackage{amsthm}
\usepackage{amsmath,mathrsfs}
\usepackage{color}
\usepackage{graphicx}
\usepackage{comment}
\usepackage{mathtools}
\usepackage{lineno}
\usepackage{algorithm}
\usepackage{algorithmicx}
\usepackage{algpseudocode}

\usepackage{amsmath}
\usepackage{graphicx}
\usepackage{natbib}

\usepackage{hyperref}
\usepackage{cleveref}

\makeatletter
\def\ps@pprintTitle{%
 \let\@oddhead\@empty
 \let\@evenhead\@empty
 \def\@oddfoot{\footnotesize\itshape
       {}}%
 \let\@evenfoot\@oddfoot}
\makeatother

\date{}

\usepackage{setspace}
\begin{document}

\begin{frontmatter}

\title{\textbf{Digital twin with automatic disturbance detection for real-time optimization of a semi-autogenous grinding (SAG) mill}}

\author[inst3]{Paulina Quintanilla}
\author[inst1]{Francisco Fern\'andez}
\author[inst1,inst2]{Cristobal Mancilla}
\author[inst1,inst2]{Mat\'ias Rojas}
\author[inst2]{Mauricio Estrada}
\author[inst1]{Daniel Navia\corref{mycorrespondingauthor}}
\cortext[mycorrespondingauthor]{Corresponding author}
\ead{daniel.navia@usm.cl}

\affiliation[inst3]{organization={Department of Chemical Engineering, Brunel University London}, country={United Kingdom}}
\affiliation[inst1]{organization={Departamento de Ingenier\'ia Qu\'imica y Ambiental, Universidad T\'ecnica Federico Santa Mar\'ia}, city={Santiago}, country={Chile}}
\affiliation[inst2]{organization={Advanced Systems, SGS Minerals}, city={Santiago}, country={Chile}}

\begin{abstract}
This work describes the development and validation of a digital twin for a semi-autogenous grinding (SAG) mill controlled by an expert system. The digital twin consists of three modules emulating a closed-loop system: fuzzy logic for the expert control, a state-space model for regulatory control, and a recurrent neural network for the SAG mill process. The model was trained with 68 hours of data and validated with 8 hours of test data. It predicts the mill's behavior within a 2.5-minute horizon with a 30-second sampling time. The disturbance detection evaluates the need for retraining, and the digital twin shows promise for supervising the SAG mill with the expert control system. Future work will focus on integrating this digital twin into real-time optimization strategies with industrial validation.

\end{abstract}


\end{frontmatter}

\section{Introduction}
\label{sec: introduction}

The rapid advancements in digital technologies have significantly transformed various industrial processes, introducing innovative methods to control and monitor operations. One notable advancement is the development of digital twins, which have gained substantial traction in recent years. A digital twin is a virtual representation of a physical system that can simulate, predict, and optimize its performance in real time \cite{MELESSE2020267}. 

Despite the increasing interest in digital twins, their application in Semi-Autogenous Grinding (SAG) mill operations still needs to be explored \cite{GHASEMI2024108733}. SAG mills are crucial in the comminution process, grinding large rocks into smaller particles for subsequent processing stages. The efficiency and stability of SAG mills directly impact the overall performance of mineral processing plants. Given the high operational costs and energy consumption associated with SAG mills, ensuring optimal performance is crucial for the economic viability of mining operations. Traditionally, expert control systems have been used to control and stabilize SAG mill operations by leveraging predefined rules and real-time data; however, their reactive nature poses significant limitations as they often respond to disturbances rather than preventing them, potentially leading to instability and inefficiencies.

The key gap in the current literature is the lack of a comprehensive digital twin framework that integrates process simulation, regulatory control, and expert control systems specifically for SAG mills. This paper integrates machine learning algorithms into the digital twin to detect anomalies and disturbances. The methodology describes the architecture of the model and the learning system. The model is trained using historical operational data from 68 hours of operation at a Chilean mining company and validated using data from 8 hours of operation. The digital twin is based on a statistical framework that predicts critical variables like bearing pressure and motor power based on inputs such as feed rate, motor speed, and solids concentration. The results and discussion section summarizes the predictive capacity of the model. Additionally, the robustness of the model and learning system is tested by simulating multiplicative disturbances in the controlled variables. 

\section{Methodology}
\label{sec: methodology}

\subsection{Model components}

The use of expert control systems in SAG milling processes stabilizes their operation. However, expert control systems are only able to forecast the long-term consequences of proposed actions within their initial design parameters. This reactive nature can result in instability problems when there are significant changes in feed conditions. For instance, alterations in feed hardness or granulometry can cause the mill to either empty or overload. In such cases, the expert control systems can only address the issue once it has been detected.

\begin{figure}
    \centering
    \includegraphics[width=0.8\textwidth]{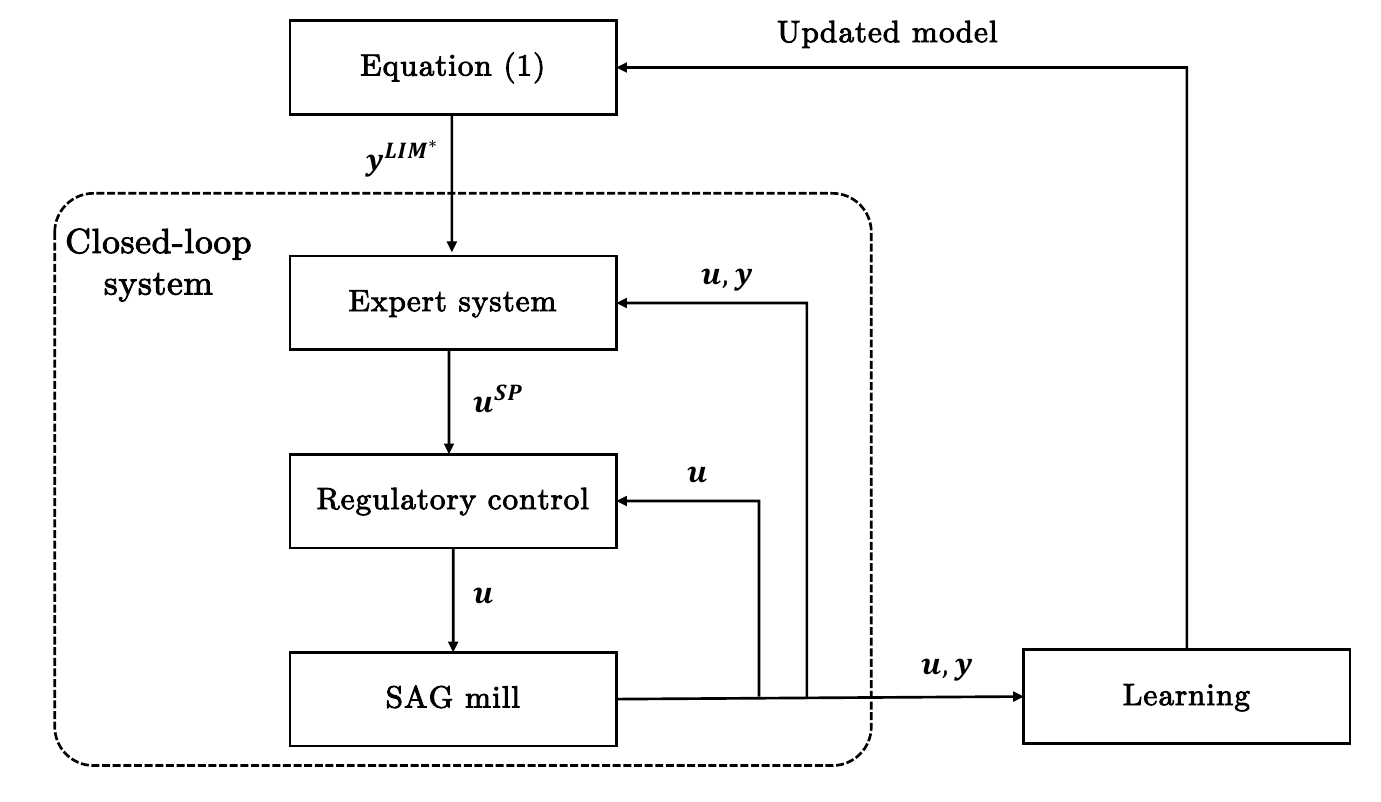}
    \caption{Proposed supervisory system scheme including a closed-loop system, where $\mathbf{u}:=[u_{1}, u_{2}, u_{3}]^T, \mathbf{u^{SP}}:= [u_{1}^{SP}, u_{2}^{SP}, u_{3}^{SP}]^T, \mathbf{y^{LIM^{*}}}:=[y_{1}^{LIM^{*}},y_{2}^{LIM^{*}}]^T, \mathbf{y}:=[y_{1},y_{2}]^T$.}
    \label{fig: system}
\end{figure}

An expert control system for SAG milling processes depends on operational conditions defined by the operational limits of the controlled variables (CV): bearing pressure ($y_1$) and motor power ($y_2$) of the mill, which can be externally modified. The studied system is shown in \Cref{fig: system}, and it comprises the following modules: 
\begin{enumerate}
\item \textbf{Expert control system}: Upper hierarchical layer with an algorithm that uses the current and past values of controlled and manipulated variables (MV) to identify the current mode of SAG operation and propose changes in the MV setpoints ($u^{SP}$).
\item \textbf{Regulatory control}: Lower hierarchical layer responsible for implementing changes proposed by the expert control system by modifying the MV values: tonnage ($u_1$), solids percentage ($u_2$), and mill speed ($u_3$) to reach their setpoint values.
\item \textbf{SAG mill:} It is the actual process to be optimized, which was simulated using recurrent neural networks as described below. 
\end{enumerate}

From \Cref{fig: system}, it can be seen that $y^{LIM}$ represents the degrees of freedom available in the closed-loop system. This implies that it is possible to implement a supervisory system over the expert control system to determine the value of $y^{LIM}$ to optimize the expected dynamic response by solving the problem in  \Cref{eq: ylim-opt} and implementing the results using a moving horizon strategy.

\begin{equation}
\label{eq: ylim-opt}
\begin{aligned}
& \min _{\boldsymbol{y}^{L I M} \in \mathcal{Y}^{L I M}} f\left(\widehat{\boldsymbol{Y}}_{\boldsymbol{k}, \boldsymbol{N}}, \widehat{\boldsymbol{U}}_{\boldsymbol{k}, \boldsymbol{N}}, \widehat{\boldsymbol{U}}_{\boldsymbol{k}, \boldsymbol{N}}^{S P}\right) \\
& \text { s.t.: } \\
& \widehat{\boldsymbol{y}}_{\boldsymbol{k} \mid \boldsymbol{k}}=\boldsymbol{h}\left(\boldsymbol{Y}_{\boldsymbol{k}, \boldsymbol{m}}, \widehat{\boldsymbol{U}}_{\boldsymbol{k}, 0}, \boldsymbol{U}_{\boldsymbol{k}, n}, \widehat{\boldsymbol{U}}_{\boldsymbol{k}, 0}^{S P}\right) \\
& \widehat{\boldsymbol{y}}_{\boldsymbol{k}+i \mid \boldsymbol{k}}=\boldsymbol{h}\left(\widehat{\boldsymbol{Y}}_{\boldsymbol{k}, \boldsymbol{i}-1}, \boldsymbol{Y}_{\boldsymbol{k}, \boldsymbol{m}-\boldsymbol{i}}, \widehat{\boldsymbol{U}}_{\boldsymbol{k}, \boldsymbol{i}}, \boldsymbol{U}_{\boldsymbol{k}, n-\boldsymbol{i}}, \widehat{\boldsymbol{U}}_{\boldsymbol{k}, \boldsymbol{i}}^{S P}\right), \quad i=1 \ldots N \\
& \widehat{\boldsymbol{Y}}_{\boldsymbol{k}, \boldsymbol{i}}:=\left[\widehat{\boldsymbol{y}}_{\boldsymbol{k} \mid \boldsymbol{k}}, \ldots, \widehat{\boldsymbol{y}}_{\boldsymbol{k}+i \mid k}\right], \widehat{\boldsymbol{U}}_{\boldsymbol{k}, \boldsymbol{i}}:=\left[\widehat{\boldsymbol{u}}_{\boldsymbol{k} \mid \boldsymbol{k}}, \ldots, \widehat{\boldsymbol{u}}_{\boldsymbol{k}+i \mid k}\right], \widehat{\boldsymbol{U}}_{\boldsymbol{k}, \boldsymbol{i}}^{S P}:=\left[\widehat{\boldsymbol{u}}_{\boldsymbol{k} \mid \boldsymbol{k}}^{S P}, \ldots, \widehat{\boldsymbol{u}}_{\boldsymbol{k}+i \mid k}^{S P}\right], \quad i=0 \ldots N \\
& \boldsymbol{Y}_{k, j}:=\left[\boldsymbol{y}_{k-1}, \ldots, \boldsymbol{y}_{k-j}\right], \quad j=1, \ldots, m, \quad \boldsymbol{U}_{k, j}:=\left[\boldsymbol{u}_{k-1}, \ldots, \boldsymbol{u}_{k-j}\right], \quad j=1, \ldots, n \\
& \widehat{\boldsymbol{y}}_{\boldsymbol{k}+i \mid k} \in \mathcal{Y}, \widehat{\boldsymbol{u}}_{k+i \mid k} \in \mathcal{U}, \quad i=0 \ldots N 
\end{aligned}
\end{equation}
\\
\noindent $k$ represents the current instant, $\mathbf{y}_{k-i}$ and $\textbf{u}_{k-i}$ correspond to the measured CV and MV at $i$ previous instants relative to $k$, and $\boldsymbol{\hat{y}}_{k+i|k}$, $\boldsymbol{\hat{u}}_{k+i|k}$, and $\boldsymbol{\hat{u}}_{k+i|k}^{SP}$ denote the corresponding variables, predicted $i$ instants into the future relative to $k$. The digital twin of the process is represented by $\textbf{h}$, which can be understood as a discrete model that allows predicting the behavior of the system's variables in the closed-loop of \Cref{fig: system}, while $f$ is the objective function of the process to optimize. In \Cref{eq: ylim-opt}, $\mathbf{\mathcal{Y}}^{LIM}$, $\mathbf{\mathcal{Y}}$, and $\mathbf{\mathcal{U}}$ describe the feasible region of the optimization problem, while $n$ and $m$ correspond to the previous instants required by the dynamic model to simulate the next value. $N$ corresponds to the prediction horizon.

\begin{figure}[h]
    \centering
    \includegraphics[width=1\textwidth]{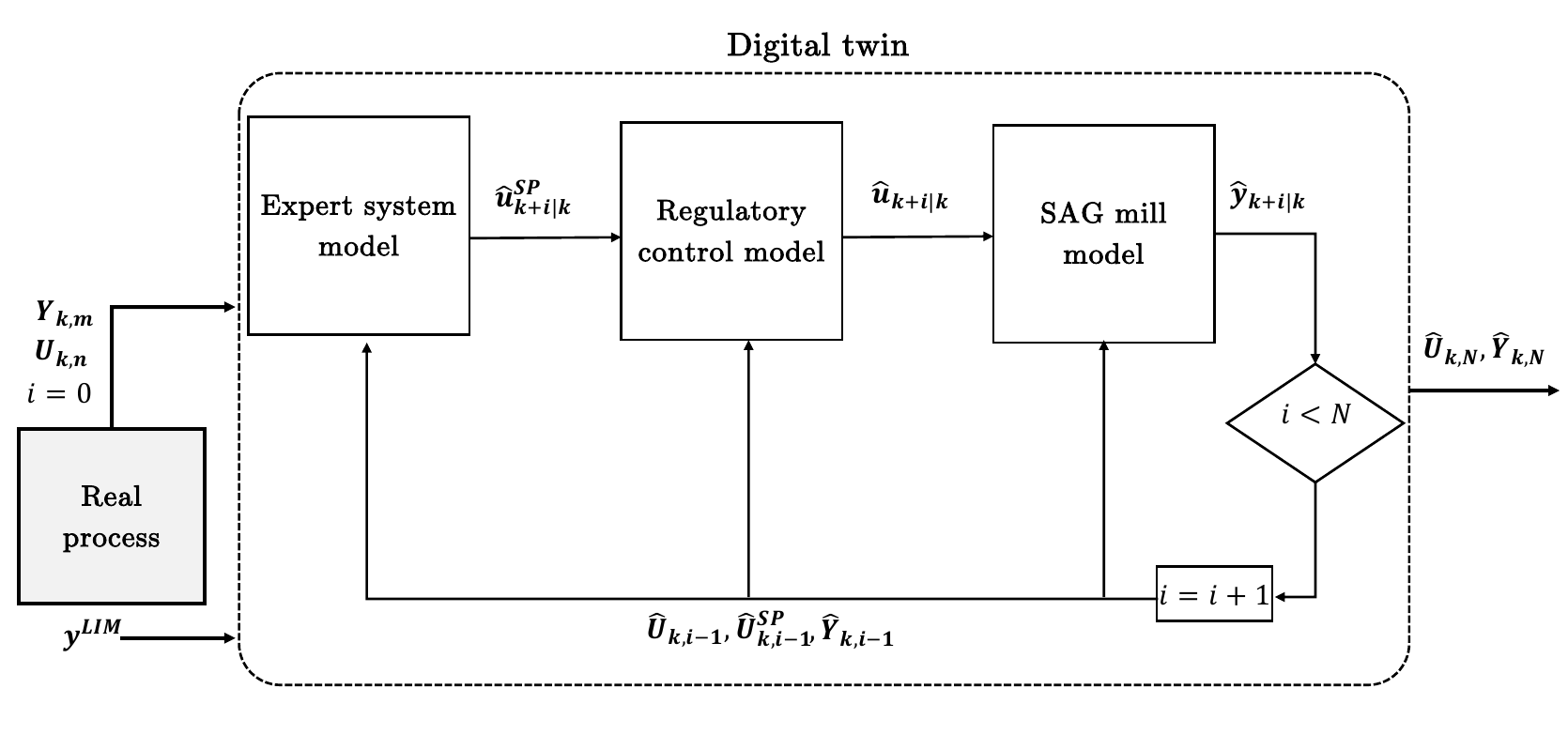}
    \caption{Digital twin structure.}
    \label{fig: fig2}
\end{figure}

The implementation of the supervisory system requires adaptability or learning capability. Learning consists of the systematic update of the process model to avoid degradation in its predictive capacity, associated with the inherent mismatch related to the assumptions made to obtain said model \cite{darby2011rto}. The learning process can be implemented by retraining the model so that it adequately predicts a window of recent data. However, overfitting should be avoided as it could lead to learning behavior associated with measurement noise. 
The digital twin emulates the architecture of the process in \Cref{fig: system} and was implemented in Python 3.7. It is composed of three modules connected in series with models of each of the sections of the closed-loop system (see \Cref{fig: fig2}).

\begin{figure}[h]
    \centering
    \includegraphics[width=1\textwidth]{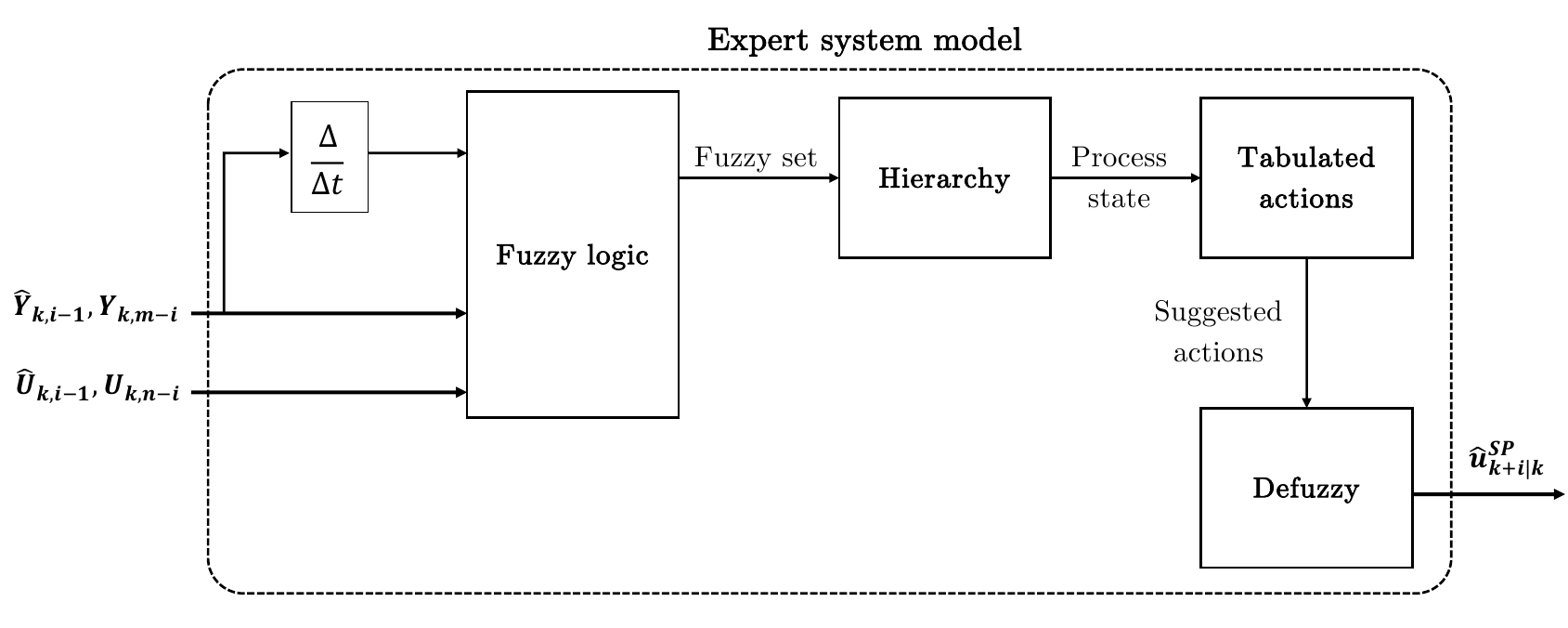}
    \caption{expert control system model description.}
    \label{fig: fig3}
\end{figure}

 $\blacktriangleright$ \textbf{Expert control system}: The expert control system model is summarized in \Cref{fig: fig3}. In the Fuzzy block, the values of the CVs and MVs, along with their slope, are transformed into fuzzy sets using triangular membership functions. The maximum of each fuzzy set determines the operating state of the process, which is associated with a hierarchy according to its criticality, and with this, the actions to be implemented in the MVs are determined. In the De-Fuzzy block, the magnitude of these actions is transformed into changes in the MVs using Sugeno \cite{sugeno1985industrial}.

$\blacktriangleright$ \textbf{Regulatory control:} The objective of the regulatory control model is to emulate the dynamic response of the control loops of the MVs. A state-space model was used, which corresponds to the discrete representation of a linear system of algebraic-differential equations (see \Cref{eq: eq2}).

\begin{equation}
\begin{aligned}
& \boldsymbol{\hat{x}}_{k+i+1 \mid k}=\mathbf{A} \boldsymbol{\hat{x}}_{k+i \mid k}+\mathbf{B} \boldsymbol{\hat{u}}_{k+i \mid k}^{S P}+\mathbf{K} \boldsymbol{e_k}, \quad \boldsymbol{\hat{x}}_{k \mid k}=\boldsymbol{\hat{x}}_0 \\
& \boldsymbol{\hat{u}}_{k+i+1 \mid k}=\mathbf{C} \boldsymbol{\hat{x}}_{k+i \mid k}+\mathbf{B} \boldsymbol{\hat{u}}_{k+i \mid k}^{S P}+\boldsymbol{e_k},
\label{eq: eq2}
\end{aligned}
\end{equation}

\noindent where $\boldsymbol{\hat{x}}_{k+i|k}$ corresponds to the state vector of the model predicted $i$ instants into the future relative to $k$, while $\mathbf{A}$, $\mathbf{B}$, $\mathbf{C}$, and $\mathbf{K}$ are matrices containing the model parameters. In \Cref{eq: eq2}, $\mathbf{e_k}$ represents a constant additive disturbance vector, and $\boldsymbol{\hat{x}}_0$ is the initial value of the states.

$\blacktriangleright$ \textbf{SAG mill model: }The SAG mill model predicts the dynamic response of the controlled variables to the changes simulated by the regulatory control model. To account for the effects of the system's free and forced response, a nonlinear autoregressive exogenous (NARX) model based on neural networks \cite{hopfield1982neural} was used, incorporating dependencies on the past values of the MVs and CVs. The dependency of the implemented NARX model is summarized in \Cref{eq:3}, where $g$ represents the model composed of the interconnection of the neural network layers.

\begin{equation}
\label{eq:3}
\widehat{\boldsymbol{y}}_{\boldsymbol{k}+\boldsymbol{i} \mid \boldsymbol{k}}=\boldsymbol{g}\left(\widehat{\boldsymbol{Y}}_{\boldsymbol{k}, \boldsymbol{i}-1}, \boldsymbol{Y}_{\boldsymbol{k}, \boldsymbol{m}-\boldsymbol{i}}, \widehat{\boldsymbol{U}}_{\boldsymbol{k}, \boldsymbol{i}}, \boldsymbol{U}_{\boldsymbol{k}, \boldsymbol{n - i} \boldsymbol{i}}\right), \quad i=0 \ldots N
\end{equation}

\subsection{Model identification and training}
The data used for parameter estimation of the models in \Cref{eq: eq2} and \Cref{eq:3} were extracted from 6 months of historical operation of the SAG, obtained every 5 seconds. This database was filtered using MySQL according to the following criteria: SAG in operation, feed above the operational minimum, percentage of solids above the operational minimum, expert control system online. The resulting data sets were filtered using a moving median between 6 contiguous data points, obtaining data sets with a sampling time of 30 seconds. The training data were selected from the largest resulting set, with 8156 time instants, equivalent to 68 hours of operation. The next largest data set (1013 time instants, equivalent to 8 hours of operation) was defined as the test data set and was used for the validation of the trained digital twin.

$\blacktriangleright$ \textbf{Regulatory control identification:} The identification of the regulatory control model is summarized in \Cref{eq:4}, where the sum of the objective function and the model applies to the entire training data horizon ($N_H$).

\begin{equation}
\label{eq:4}
\begin{aligned}
& \min _{A, B, C, K, \widehat{x}_0} \sum_{k=1 \ldots N_H}\left\|\boldsymbol{u}_{\boldsymbol{k}}-\widehat{\boldsymbol{u}}_{\boldsymbol{k} \mid \boldsymbol{k}}\right\|_2 \\
& \text { s.t.: Eq. (2), with } \boldsymbol{e}_{\boldsymbol{k}}=\mathbf{0}
\end{aligned}
\end{equation}

To solve \Cref{eq:4}, the order of the model must be defined, which can be interpreted as the number of previous instants required for the prediction. To determine this value, the following procedure was implemented: 1) set the value of the model order, 2) estimate parameters. This procedure is performed for different model orders, and the one with the smallest order, whose fit does not significantly improve compared to higher orders, is selected.

$\blacktriangleright$ \textbf{SAG mill model identification: }The identification of the SAG mill model is summarized in \Cref{eq:5}, where the sum of the objective function and the model applies to the entire training data horizon. NARX summarizes the parameters of the neurons: input weights, bias, and activation. Only one hidden layer was used.

\begin{equation}
\label{eq:5}
\begin{aligned}
& \min _{\text {NARX }} \sum_{k=1 \ldots N_H}\left\|\boldsymbol{y}_{\boldsymbol{k}}-\widehat{\boldsymbol{y}}_{\boldsymbol{k} \mid \boldsymbol{k}}\right\|_2 \\
& \text { s.t. : Eq. (3) }
\end{aligned}
\end{equation}

To solve \Cref{eq:5}, the number of neurons in the hidden layer and previous instants of the MVs and CVs must be defined. The following procedure was implemented for their estimation: 1) set the number of previous instants and neurons in the hidden layer, 2) estimate parameters. This procedure is performed for different combinations of previous instants and the number of neurons, and the combination with the smallest number of previous instants and hidden neurons, whose fit does not significantly improve compared to models with larger combinations, is selected.

\subsection{Automatic disturbance detection}
The automatic disturbance detection acts as a learning module to update the models of the digital twin for regulatory control and the SAG mill. The update of the expert control system model is not considered since it is subject to changes in the real expert control system, a condition controlled externally. The learning of the regulatory control model consists of estimating $\boldsymbol{\hat{x}}_0$ and $\mathbf{e}_k$ by solving \Cref{eq:6} each time the digital twin is used (it is assumed at each instant $k$). The estimation horizon ($N_E$) corresponds to the number of previous instants used for the estimation. In \Cref{eq:6}, $ \boldsymbol{A}^*$, $ \boldsymbol{B}^*$, $ \boldsymbol{C}^*$, and $ \boldsymbol{K}^*$ represent the results of \Cref{eq:4}.

\begin{equation}
\label{eq:6}
\begin{aligned}
& \min _{\boldsymbol{\widehat{x}}_0, \mathbf{e}_k} \sum_{i=1 \ldots N_E}\left\|\boldsymbol{u}_{\boldsymbol{k}-\boldsymbol{i}}-\widehat{\boldsymbol{u}}_{\boldsymbol{k}-\boldsymbol{i} \mid \boldsymbol{k}}\right\|_2 \\
& \text { s.t. : Eq. (2) with } \boldsymbol{A}^*, \boldsymbol{B}^*, \boldsymbol{C}^*, \boldsymbol{K}^*
\end{aligned}
\end{equation}

\begin{figure}[h]
    \centering
    \includegraphics[width=1\textwidth]{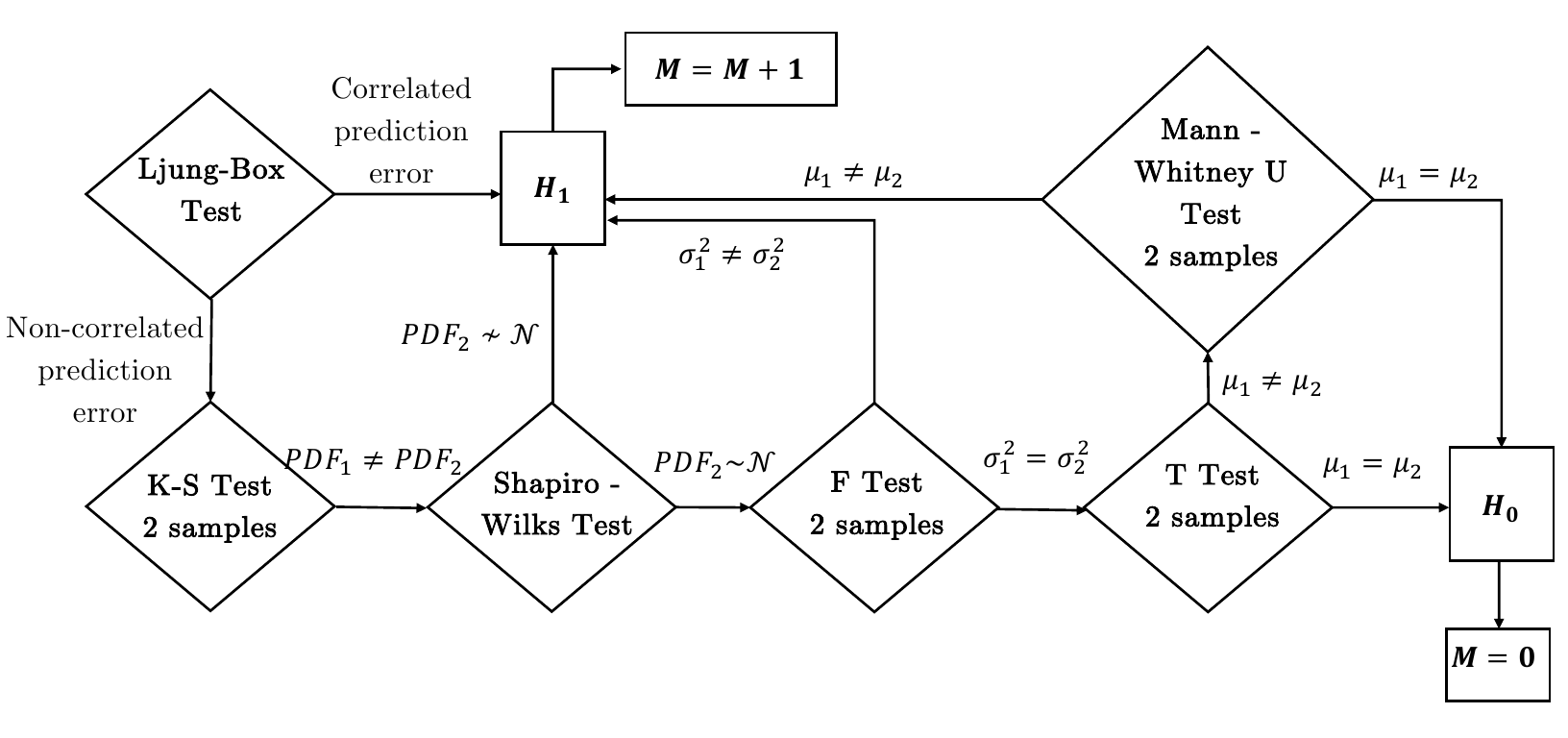}
    \caption{Diagram of the learning detection algorithm}
    \label{fig: fig5}
\end{figure}

\begin{figure}[h]
    \centering
    \includegraphics[width=1\textwidth]{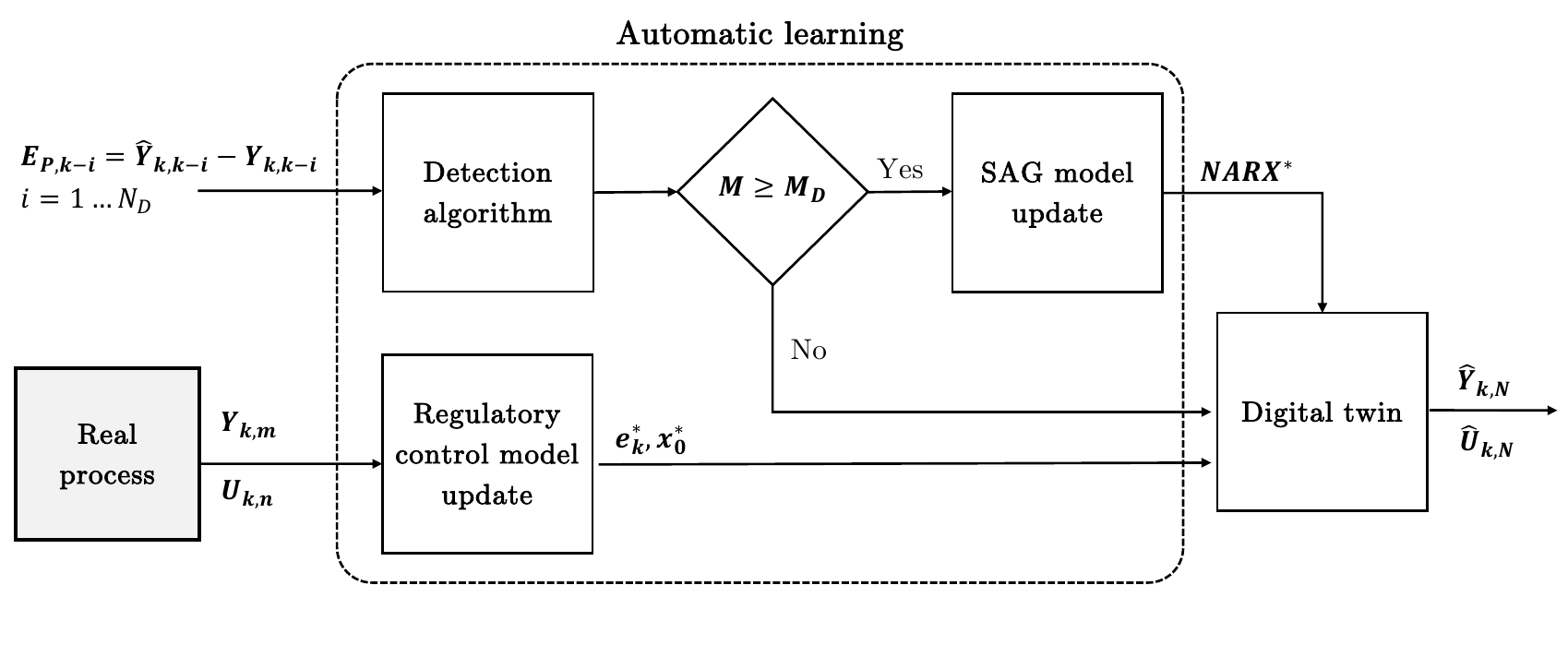}
    \caption{Diagram of the digital twin integrated with the learning detection algorithm.}
    \label{fig: fig6}
\end{figure}

The learning of the SAG model consists of two parts: detection and retraining. The idea of implementing the detection module is to avoid overfitting with measurement noise since neural networks are universal regressors. The detection module (\Cref{fig: fig5}) evaluates whether the model prediction deteriorates significantly compared to the training results at each instant $k$. For this, the characteristics of the prediction error between two data sets are compared: (1) training and (2) the one obtained in a window of $N_D$ recent data points. This comparison is carried out through two-sided hypothesis tests for error correlation, probability distribution (PDF), variance, and mean. If no null hypothesis is rejected, the counter $M=0$ is set; otherwise, the counter is updated $M=M+1$. This procedure is performed at each sampling instant. If the value of $M$ exceeds the detection threshold $M_D$, then the SAG model is retrained by solving \Cref{eq:5} with the most recent $N_E$ process data that meet the training criteria. \Cref{fig: fig6} describes the digital twin scheme with the machine learning system.

\section{Results and discussions}
\label{sec: results}

\subsection{Digital twin simulation}
\begin{figure}
    \centering
    \includegraphics[width=1\textwidth]{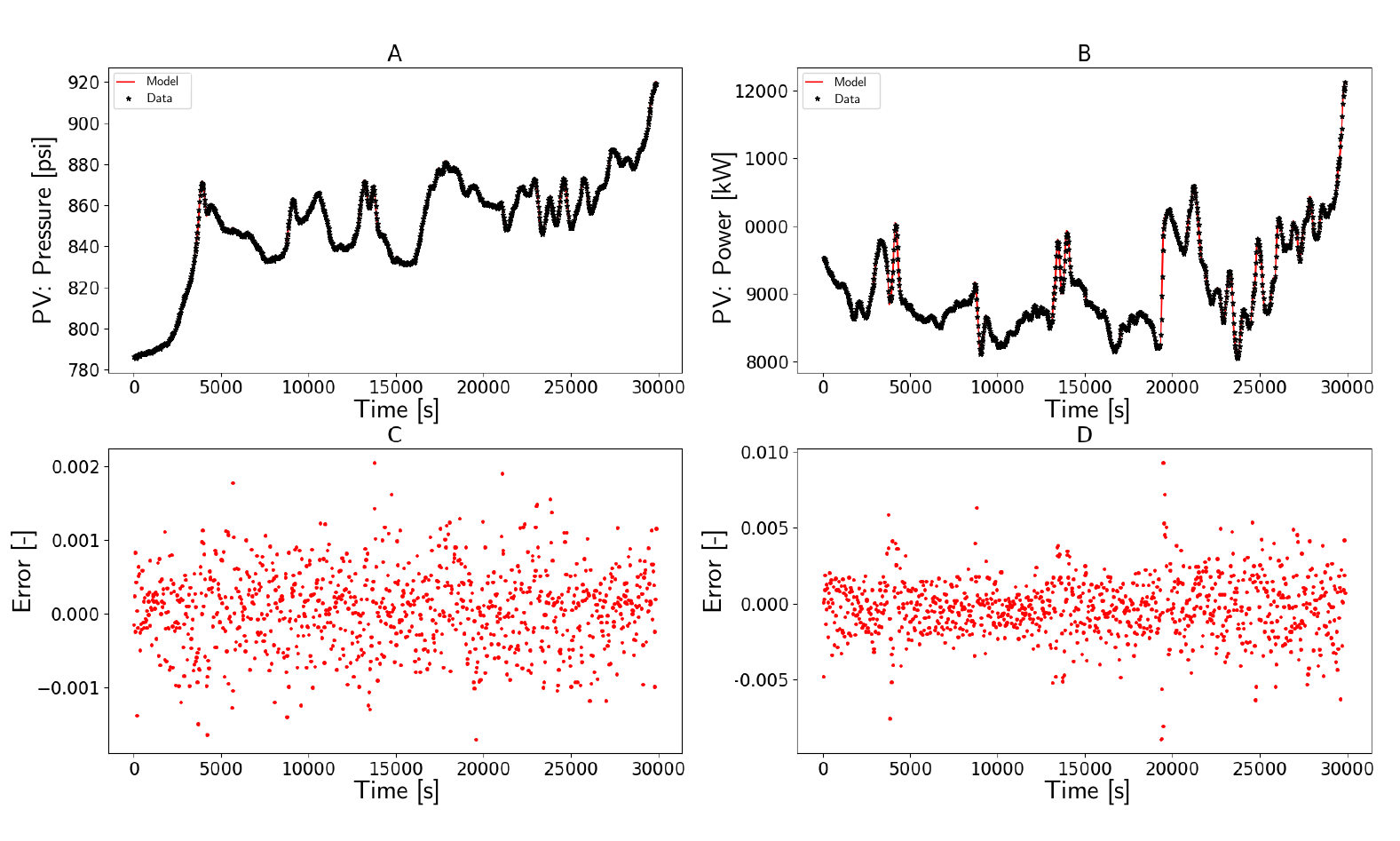}
\caption{(A) Test data and GD prediction for pressure (C) Respective proportional errors. (B) Test data and GD prediction for motor power (D) Respective proportional errors.}
    \label{fig: fig7}
\end{figure}

\begin{figure}
    \centering
    \includegraphics[width=1\textwidth]{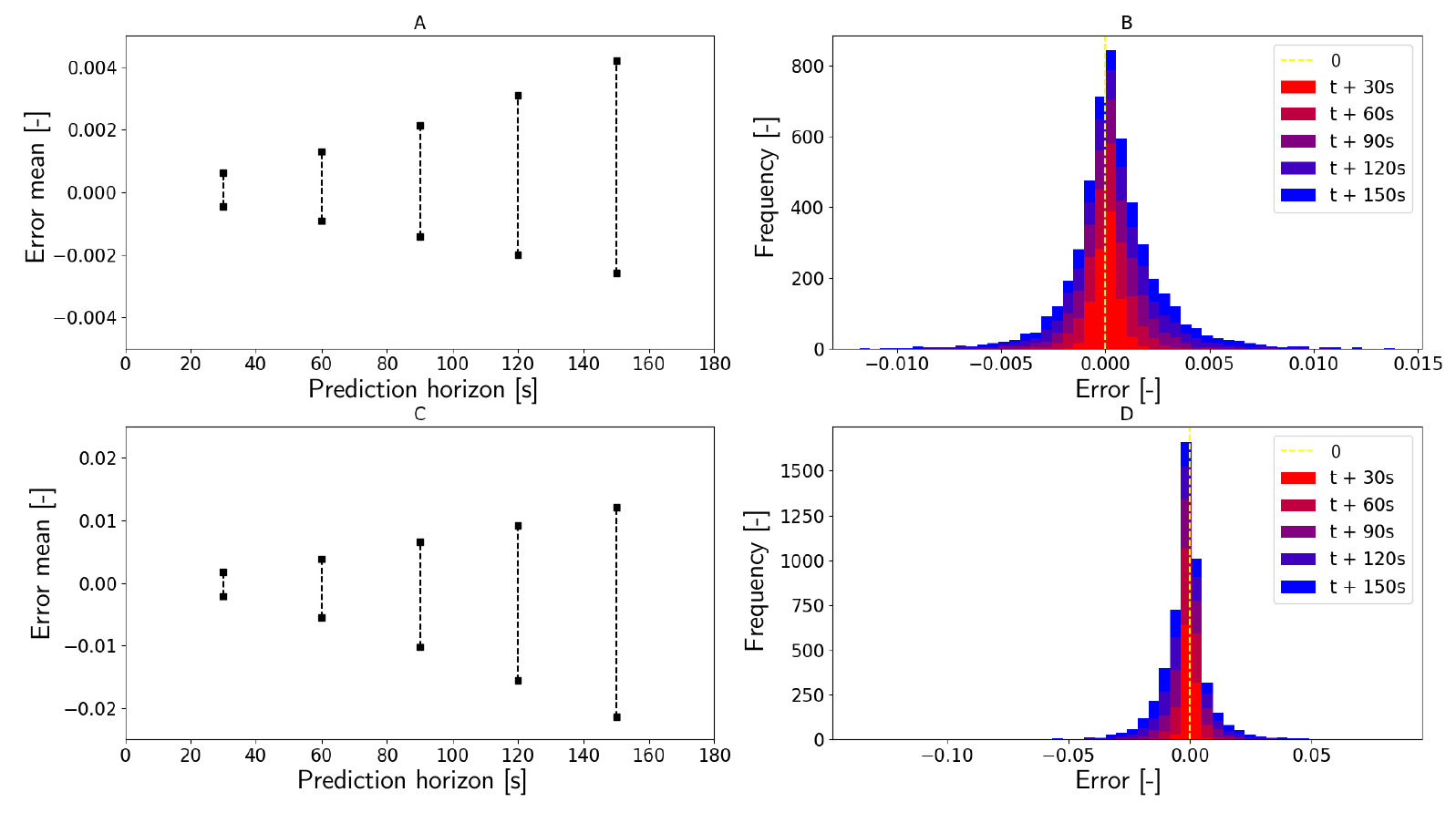}
\caption{Standard prediction error interval for (A) Pressure and (C) Power. Histograms of prediction errors for (B) Pressure, and (D) Power.}
    \label{fig: fig8}
\end{figure}

The SAG mill process was simulated using NARX models with 1 hidden layer composed of 2 neurons and 12 previous time instants were used. \Cref{fig: fig7} represents the test data and their predictions, while \Cref{fig: fig8} summarizes the main characteristics of the digital twin's prediction error for 5 prediction instants.

\Cref{fig: fig7} shows that the digital twin describes the trend of the process data. Regarding prediction errors, \Cref{fig: fig8} (A and C) shows that both the prediction of the pressure and the motor power are centered around zero, with dispersion increasing as the prediction horizon extends, due to the feedback of the simulated CV errors associated with the autoregressive characteristic of the NARX model. Based on the obtained data, the prediction error for 2.5 minutes of prediction is expected to be within the interval \([-1\%, 1\%]\) for pressure and \([-5\%, 5\%]\) for power, with 99.\% certainty. Regarding the error distribution, the histograms in \Cref{fig: fig8} (B and D) show a normal behavior centered around zero for all prediction instants. Considering also that the data showed no correlation with the explanatory variables, it can be indicated that the digital twin adequately describes the closed-loop system in \Cref{fig: system}.

\subsection{Automatic detection algorithm for learning}

The detection algorithm uses 30 previous time instants as the analysis window. The value $M_D$ was 103 for the pressure at the mill rests and 181 for the motor power.

The detection capability of the learning algorithm was evaluated by modifying the test data set, adding multiplicative perturbations in the controlled variables to emulate expected behavior as the mill's metal linings wear out after 1 month and 5 months of continuous operation. For this, a constant wear of 2\% per month was assumed, inversely proportional to the pressure at the rests and directly proportional to the mill's rotation speed. Similarly, the simulation data set was modified to emulate expected behavior with a 10\% increase in ore hardness, assuming this increase is directly proportional to the pressure at the rests and motor power. These scenarios are only indicative and are defined to operationally relate the detection system's behavior to multiplicative perturbations affecting data correlation.

\begin{figure}
    \centering
    \includegraphics[width=1\textwidth]{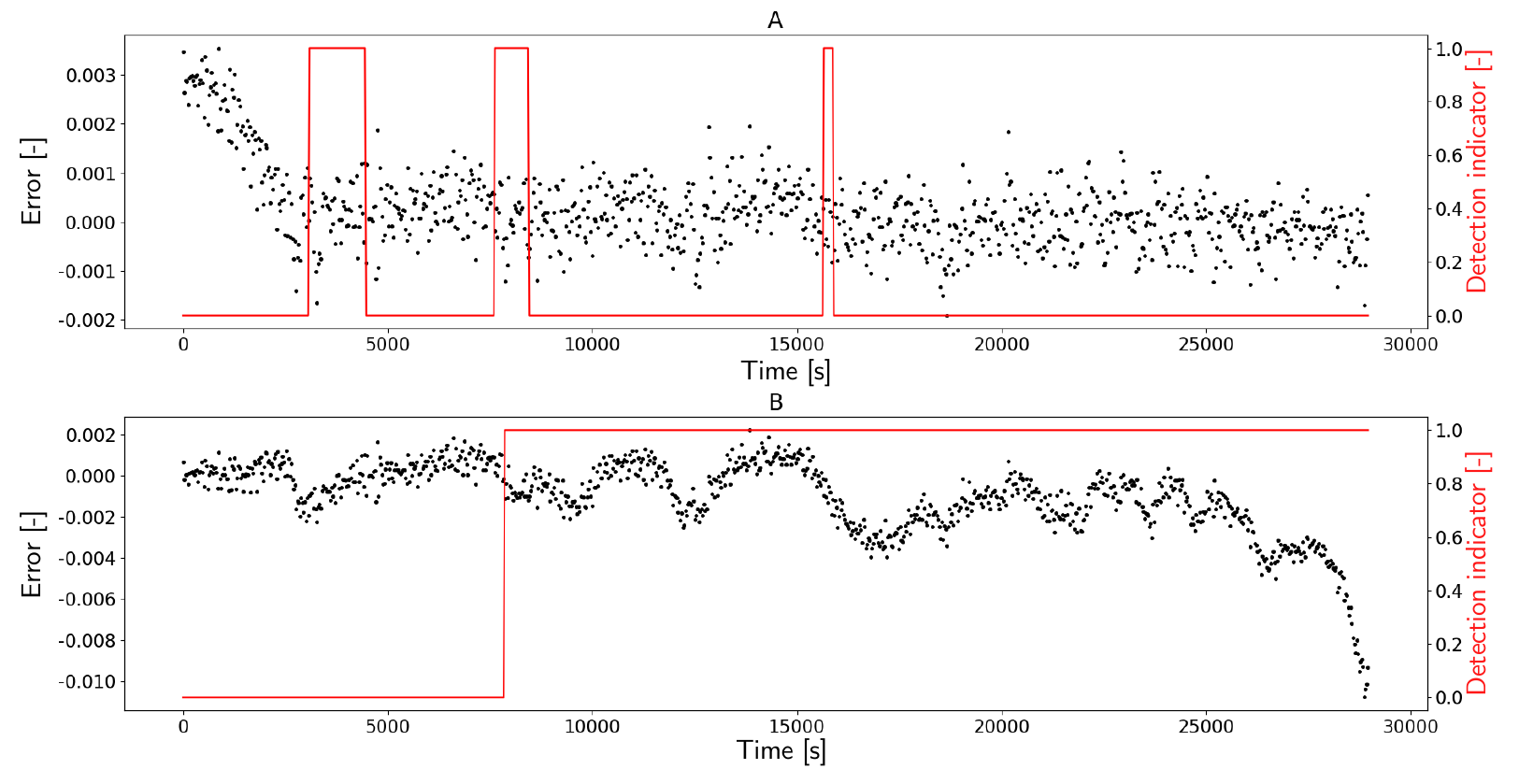}
\caption{Proportional errors of the rest pressure model and learning detection indicator for (A) mill lining wear scenario after 5 months of operation, (B) 10\% increase in ore hardness.}
    \label{fig: fig9}
\end{figure}

For the evaluated scenarios, the algorithm did not detect a need for training for the motor power. In the case of the pressure model, the detection system did not identify a need for updating for the 1-month lining wear scenario. For the 5-month wear and hardness increase scenarios, the training detection system detected a need for updating, as shown in \Cref{fig: fig9}. In the simulated 5-month wear scenario, the detector indicates the need for training at three instants in the modified data set (\Cref{fig: fig9}A). These indicators are triggered because the model consistently underestimates the pressure value, implying that the mean prediction error becomes less than zero, a consequence of the mismatch associated with the multiplicative perturbation that increases the process pressure value by 10\%. For the hardness increase scenario (\Cref{fig: fig9}B), the training detector activates from 7000s onward, due to the prediction error correlation, associated with the loss of the ability to describe the trend of the SAG's measured data.

\section{Conclusions}
\label{sec: conclusions}
Based on the obtained results, it is concluded that the implemented digital twin predicts the behavior of the controlled variables of the SAG mill operated with an expert control system, with a maximum prediction error of 5\% for 2.5 minutes and less than 1\% for a 30-second prediction. Considering that this model will be implemented in a supervisory system that follows a moving horizon strategy, it can be concluded that the digital twin is suitable for providing predictive capabilities to the current supervisory module.

Regarding automatic disturbance detection, it can be concluded that the training detector is capable of identifying conditions in which the digital twin loses the model's predictive ability due to changes in the correlation between the process's input and output variables, thus avoiding overfitting. This feature indicates that the supervisory system based on the RTO paradigm to be implemented would not significantly degrade its performance due to expected changes in process operation, as the detection module would identify this situation, and the retraining procedure would adjust the model to describe the current state of the process.

 \bibliographystyle{elsarticle-num} 
 \bibliography{cas-refs}

\end{document}